\documentstyle[prl,aps,epsf]{revtex}

\newcommand{\beq}{\begin{equation}}
\newcommand{\beqn}{\begin{eqnarray}}
\newcommand{\eeq}{\end{equation}}
\newcommand{\eeqn}{\end{eqnarray}}
\newcommand{\vp}{\varphi}

\newcommand{\ts}{  \textstyle}

\def\({\left(}
\def\){\right)}
\def\[{\left[}
\def\]{\right]}

\def\vpz{\varphi_{0}}

\def\P{\Phi}
\def\vp{\varphi}

\def\t{\tilde}
\def\tPk{{\t{\P}_k}}

\def\tphi{\t{\phi}}
\def\tchi{\t{\chi}}
\def\tQ{\t{Q}}

\def\tQki{{\t{Q}_k^i}}
\def\tP{\t{\Phi}}

\def\tvp{\tilde{\varphi}}
\def\tvpi{{\tvp^i}}
\def\tvpj{{\tvp^j}}
\def\tdvp{\delta\t{\varphi}}
\def\tQkp{ \tilde{Q}^{\varphi}_k}
\def\tQkc{ \tilde{Q}^{\chi}_k }

\def\H{{\cal{H}}}

\def\V{{\cal{V}}}

\def\k{\kappa}

\begin{document}
\draft
\twocolumn[\hsize\textwidth\columnwidth\hsize\csname
@twocolumnfalse\endcsname

\preprint{RCG; hep-ph/9909353}

\title{{\bf Massless metric preheating}}

\author{ Bruce A.
Bassett$^{\dagger,\ddagger}$
and Fermin Viniegra$^{\dagger}$}
\address{$\dagger$ Department of Theoretical Physics,
Oxford University, Oxford~OX1~3NP, England}
\address{$\ddagger$ Relativity and Cosmology Group, School of Computer
Science and Mathematics, Portsmouth University, Portsmouth~PO1~2EG,
England}
\date{\today}
\maketitle

\begin{abstract}
Can super-Hubble metric perturbations be amplified exponentially during
preheating ? Yes. An analytical existence proof is provided by exploiting the
conformal properties of  massless inflationary models. The traditional
conserved quantity $\zeta$ is non-conserved in
many regions of parameter space. We include  backreaction through the
homogeneous parts of the inflaton and preheating fields and discuss the
r\^ole of initial conditions on the post-preheating power-spectrum.
Maximum field variances are strongly underestimated if metric
perturbations are ignored. We illustrate this in the case of strong
self-interaction of the decay products. Without metric perturbations,
preheating in this case is very
inefficient. However, metric perturbations increase the maximum
field variances and give alternative channels for the resonance to
proceed. This implies that metric perturbations can have a large impact on
calculations of  relic abundances of particles produced during preheating.
\end{abstract}
]

\section{Introduction}

Reheating is a crucial element of any inflationary cosmology.  It endows
the nascent universe with the high temperatures and large entropy needed
for its subsequent evolution.  Reheating changes the equation of state
from the near de Sitter $p \simeq - \rho$, to the radiation-dominated
form $p = \ts{1\over3} \rho$. This leads to a large amplification of
the power spectrum of super-Hubble metric perturbations resulting in a
final value $P_k \propto \dot{\phi}^{-1}|_{k = aH}$, where $\phi$ is the
slowly-rolling inflaton field \cite{MFB,KS84}. This we call the `old'
theory of cosmological reheating. It is a theory where
coherence of the inflaton, and the fact that the Hubble scale, $H^{-1}$,
at the end of inflation  is dwarfed by the vastly larger particle horizon
$d_H$,  play no r\^ole.

In contrast, the super-Hubble coherence \cite{early} of the inflaton
condensate forms the  foundation of preheating \cite{KLS94}. It is
an intrinsically non-equilibrium era \cite{ctp} in which  stimulated,
coherent processes can totally dominate single-body, perturbative decays.
The resulting phase of violent particle production leaves as its legacy a
non-thermal spectrum of particles. The subsequent
backreaction and  rescattering \cite{rescatt} of these fields leads
to turbulence and thermalisation \cite{therm}.

While many different models have been proposed, the basic mechanism of
parametric resonant amplification of quantum fluctuations is very robust
and bears many similarities to standard quantum field theory in strong
external fields \cite{dk,strongfield}.  Given this robustness, a natural
question is whether these time-dependent mass effects have implications
for the standard predictions of inflationary cosmology \footnote{It is
important to distinguish between metric amplification due (1) purely to
resonances which stem from reversible changes to effective masses and (2)
to thermalisation which involves irreversible change in
the average equation of state and large  entropy production.}.
Primary among these predictions
 is the spectrum of metric perturbations, which will provide precision
tests of the inflationary paradigm. It is therefore of great interest to
understand these issues in detail.

That parametric resonance due to oscillating scalar fields might have an
interesting impact on metric perturbations was first noted by Kodama and
Sasaki \cite{KS84}. The first works addressing this issue within the
context of preheating however claimed that there
would be little effect on metric perturbation evolution on super-Hubble
scales, and in particular that $\zeta$ (see eq. \ref{zeta}), would remain
conserved \cite{early2,later}. These conclusions stem from considering
small resonance parameters $q \leq $ O(1) such as occurs in the single
field $\lambda \phi^4$ model \cite{KLS94}. While these models can yield a
rich spectrum of sub-Hubble physics \cite{nonlin} they cannot be
considered generic models of preheating, which {\em usually} requires $q
\gg 1$ in order to be efficient \cite{rescatt}.

Detailed simulations \cite{BTKM} have presented strong evidence in support
of the intuitive arguments \footnote{Such as the fact that gravity has
{\em negative} specific heat: inhomogeneity is the natural end-point of
gravitational evolution.} for the existence of robust resonances on
super-Hubble ({\em a.k.a.} ``super-horizon") scales \cite{BKM}.  Such
amplification does not violate causality \cite{BKM,BTKM}. Nevertheless, an
{\em analytical} existence proof of the possibility of exponential growth
of super-Hubble metric perturbations remained outstanding. This {\em
Letter} aims to provide this missing link. We will show that preheating
can in principle cause significant departures from the old theory of
metric perturbation evolution through reheating.

The failure of the old theory is most glaringly  seen in the violent,
essentially exponential\footnote{By {\em essentially exponential} we
mean exponential modulo sub-dominant power-law factors; i.e. $e^{\mu
t}/t^{\alpha}$, $\alpha \leq 1$.}, growth of the Bardeen potential
$\zeta$. In the old theory, $\zeta$ is exactly conserved in the adiabatic,
long-wavelength limit. This allowed one to match the scalar metric
perturbation $\Phi$ across the desert  separating the Grand Unified  scale
and photon decoupling where temperature anisotropies in the CMB were
formed on large angles with $\Delta T/T \simeq \ts{1\over3}\Phi$ ($\Omega = 1,
\Lambda = 0$).

In section (\ref{gen}) we set out the general class of models we consider
together with the general, conformally rescaled multi-field equations.
Section (\ref{archetype}) generalises the work of \cite{GKLS97,dk2} to
include metric perturbations. This provides the basic analytical results
of our paper. Section (\ref{init}) discusses the appropriate initial
conditions for the $\chi$ field at the start of preheating while section
(\ref{power}) discusses the $\chi$ power spectrum at the start of
preheating, one of the key factors in evaluating the impact of preheating
on the CMB. Finally section (\ref{back}) discusses the effects of
backreaction and self-interaction among the decay products. We use natural
units where $G = 1$.

\section{The general class}\label{gen}

Our general potential for the class of massless scalar fields
$\vp_i$, interacting with couplings $g_{ijkl}$,
will be:
\beq
V(\vp) = \sum_{i,j,l,m}^{N} g_{ijlm}~ \vp_i \vp_j \vp_l \vp_m\,.
\label{genpot}
\eeq
This includes many interesting sub-classes such as the archetypal
$\lambda \phi^4 + g^2 \phi^2 \chi^2$ model considered in section
(\ref{archetype}). We choose
this form since it is the most general one consistent with conformal
invariance \footnote{We neglect the conformal coupling to the curvature
since the associated resonance is weak for $\xi = 1/6$
\cite{BL98}.}.
The fluctuations in these fields generate scalar metric perturbations
\footnote{We neglect vector and tensor perturbations \cite{gw} and
consider a flat FLRW background. See  \cite{dk2,VB} for aspects of the
non-flat case.} which, in the
longitudinal gauge, are encoded in the perturbed metric
\beq
 ds^2=a(\eta)^2 \[\(1 -2\P\)d\eta^2 -\(1 +2\Psi\)dx^idx_i\]\,,
\label{metric}
\eeq
where $d \eta = dt/a$ is the conformal time, $a(\eta)$ is the scale
factor, and $\Phi = \Psi$ since the anisotropic stress of
the system vanishes to linear order \cite{BTKM}. It proves very useful to
re-scale all the fields by $a(\eta)$. We adopt the convention that
\underline{$\tilde{F}\equiv aF$}  for any field $F$.
The spatially homogeneous parts of the fields satisfy the background
Klein-Gordon equations:
\beq
 \tvp_i'' - \ts{a''\over a}\tvp_i + \V_{\tvp_i}=0\,,
 \eeq
where $\prime \equiv d/d\eta$, $\V_{\tvp_i} \equiv \partial \V/\partial
\tvp_i$ and  $\V(\tvp)
\equiv a^4 V(\vp)$ is the conformal potential, given by
 \beq
 \V(\tvp) = \Sigma_{i,j,l,m}~ g_{ijlm}~ \tvp_i \tvp_j \tvp_l  \tvp_m\,.
 \eeq
The perturbed, multi-field, Einstein equations are:
\beqn
\tdvp^{i \prime\prime}_k + k^2\tdvp^i_k &+& \sum_j
\V_{\tvpi\tvpj}\tdvp^j_k
= \frac{4}{a}\[ \tilde{\Phi}'_k \tvp^{i\prime} -
\frac{1}{2}\V_{\tvpi}\tilde{\Phi}_k\]
\nonumber \\
&-&\frac{4a'}{a^2}\[\frac{\tP\tvp}{a}\]'\,,
\label{bigphi} \\
\tilde{\Phi}_k' &=& \frac{4\pi}{a}~ \sum_i (\tvp^i{}' - \H
\tvp^i)\tdvp_k^i\,.
\label{constraint}
\eeqn
We now reformulate this system in terms of the rescaled Sasaki-Mukhanov
variables:
\beq
\tQki \equiv \tdvp_k^i + \frac{\tvp^i{}'}{a'}\tP_k -
\frac{\tvp^i}{a}\tP_k\,,
\label{smvar}
\eeq
which satisfy the equations:
$$
\tQ_k^i{}'' + \[k^2 - \frac{a''}{a}\]\tQki
 + \sum_{j}\V_{\tvpi\tvpj}\t{Q}_k^j  =
\frac{8\pi}{a}\sum_{j}\tilde{M}^{ij}\t{Q}_k^j \,.
$$
The constraint equation (\ref{constraint}) can be recast as:
\beq
\tilde{\Phi}'_k + \[2\H-\frac{a''}{a'}\]\tilde{\Phi}_k =
\frac{4\pi}{a} \sum_i(\tvpi{}' -
\H\tvpi) \tQki\,,
\label{constraint2}
\eeq
where $\H=a'/a$. Note that when $a'' =  0$, the LHS of these equations are
expansion-invariant (EI) while the non-EI coefficients on the RHS
are:
\beqn
\tilde{M}^{ij} &\equiv&
\[\frac{\tvpi{}' \tvpj{}'}{a'}\]'+
\frac{1}{a}\[\tvpi{}'\tvpj{}'-\(\tvpi\tvpj\)''\]
\nonumber \\
&+&\frac{1}{a^2}\[a'\tvpi\tvpj\]'
 -\frac{a'{}^2}{a^3}\[\tvpi\tvpj\]\,.
\label{mij}
\eeqn
Finally, the traditional ``conserved" quantity $\zeta$, rescaled by $a$,
is given
by:
\beq
\tilde{\zeta}_k = \tPk + \[2-\frac{a''}{a'}\]^{-1}\H^{-1}\tPk'\,.
\label{zeta}
\eeq
Equations (\ref{smvar}-\ref{zeta}) will provide the foundation for our
analytical and numerical results.
\section{An archetypal example}\label{archetype}

Within the general class of potentials (\ref{genpot}), a key,
well-studied  example is the massless $\lambda \phi^4$ model coupled
quartically to a massless $\chi$ \cite{GKLS97,dk2}. The
conformal potential in this case is
\beq
\V(\tphi,\tchi) =  \frac{\lambda}{4} \tphi^4 + \frac{g^2}{2}\tphi^2
\tchi^2 \,.
\label{2fieldpot}
\eeq
We switch to dimensionless time  $x \equiv \sqrt{\lambda} \vpz \eta$ where
$\varphi_0 \simeq 0.3$ is the value of $\tphi$ at the start of preheating
corresponding to  $x_0 \simeq 2.44$ \cite{GKLS97}.  In this model, the
time-averaged equation of state is that of
radiation and the scale factor obeys:
\beq
a \simeq  (\ts{2\pi\lambda\over3})^{1/2}~\vpz^2 \eta =
(\ts{2\pi\over3})^{1/2}~\vpz x\,.
\eeq
This implies that $a'' \simeq 0$. Define $\tilde{\varphi}
\equiv \tilde{\phi}/\varphi_0$. Then  $\tilde{\varphi}$ satisfies:
\beq
\tvp'' + \tvp^3  + \ts{g^2\over \lambda\varphi_0} \tilde{\chi}^2\tvp =0\,,
\label{inflaton}
\eeq
which during the first stage of preheating when $\tilde{\chi} \simeq 0$,
has the oscillating solution \cite{ctp,GKLS97} $\tilde{\varphi}(x)= cn(
x-x_0,1/\sqrt{2})$ where $cn$ is the Jacobi elliptic cosine,  while
\beq
\tchi'' + \frac{g^2}{\lambda}\tvp^2\tchi =0\,.
\label{tc}
\eeq
The  $\tQki$ satisfy:
\beqn
\tQkp{}'' &+& \[\k^2 + 3 \tvp^2\]\tQkp  =
-2\frac{g^2}{\lambda\varphi_0} \tvp \tchi \tQkc
\nonumber \\
 &+&  \frac{8\pi}{a}\( \tilde{M}^{\varphi\varphi}\tQkp +
\tilde{M}^{\varphi\chi}\tQkc\)
\label{qphi} \\
\tQkc{}'' &+& \[\k^2 +
\frac{g^2}{\lambda} \tvp^2 \]\tQkc
= -2\frac{g^2}{\lambda\varphi_0} \tvp \tchi
\tQkp  \nonumber\\ &+& \frac{8\pi}{a}\( \tilde{M}^{\varphi\chi}\tQkp +
\tilde{M}^{\chi\chi}\tQkc\)\,,
\label{qchi}
\eeqn
where $\kappa^2 \equiv k^2/\lambda\varphi_0^2$.
These equations split naturally into terms which are expansion-invariant
(EI), and those which are non-EI which come from the
$\tilde{M}^{ij}$ in eq. (\ref{mij}) and which depend explicitly on $a$.
We now consider these equations with and without $\chi$ backreaction.

\subsubsection{ Case 1: Reduction to generalized Lam\'e
form}

\begin{figure}
\epsfxsize=3.5in
\epsffile{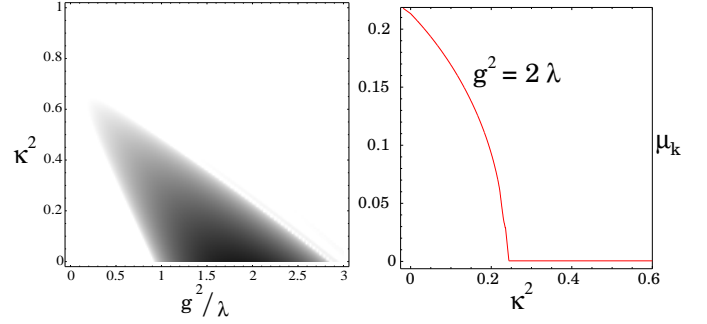}
\caption{Floquet index, $\mu_{\kappa}$, instability chart. White
represents $\mu_{\kappa} = 0$, darker greys represent  increasing
$\mu_{\kappa}$ (left). $\mu_{\kappa}$ vs $\kappa^2 \equiv
k^2/\lambda\varphi_0^2$ for $g^2/\lambda = 2$ (right). Both plots are for
the generalized Lam\'e equation given by Eq. (\ref{qchi}) with the RHS set
to zero and $\tvp = cn(x,1/\sqrt{2})$.}
\label{fig0}
\end{figure}

Let us first study the equations (\ref{inflaton}--\ref{qchi}) without any
backreaction \footnote{By backreaction we mean including terms involving
products of the background $\chi$ field and a genuine, inhomogeneous,
first order variable such as $\tQkc$. While $\tchi$ is homogeneous, so
that the product is ostensibly first order, $\tchi$ is  vanishingly small
during inflation. Hence it is numerically irrelevant at the start of
preheating. Once $\tchi$ grows large it cannot be neglected and
leads to the termination of the resonance (see Fig. \ref{fig2}).}, i.e.
using $\tvp = cn(x,1/\sqrt{2})$.
This eliminates all terms on the RHS except that containing
$\tilde{M}^{\varphi\varphi}$. We verified numerically that this term
provides only a weak, super-Hubble, {\em  growing} mode to $\tQkp$.
The system is decoupled, exact Floquet theory can be applied to the
$\tQkc$ equation and we are guaranteed solutions growing exponentially as
$e^{\mu_{\kappa} x}$ for values of $\kappa$ and
$g^2/\lambda$ inside resonance bands; see Fig. (\ref{fig0}).  Upon
neglect of $ \tilde{M}^{\varphi\varphi}$,
Floquet theory can also be applied to the $\tQkp$ equation.

\begin{figure}
\epsfysize=2.3in
\epsffile{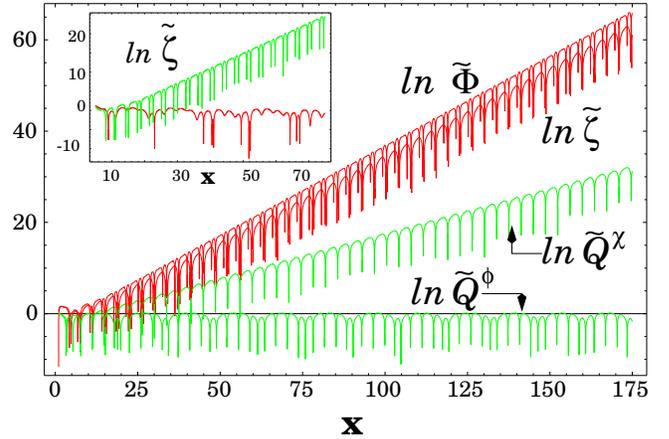}
\caption{The gauge-invariant metric perturbations $\tilde{\Phi}$ and
$\tilde{Q}^{\chi}_k$, $\tilde{Q}^{\vp}_k$ for the super-Hubble mode
$\kappa^2 = 10^{-20}$ and $g^2/\lambda = 2$. Note that $\tilde{\Phi}_k$
grows with Floquet
index $\mu \simeq 0.357$, much larger than the maximum possible in
preheating neglecting metric perturbations which is bounded for all values
of $g^2/\lambda$ to be less than or equal to $0.238$. {\bf Inset}:
$\tilde{\zeta}_k$ for $g^2/\lambda = 2$
($\tilde{\zeta}_k$ exponentially growing) and $g^2 = 0$
($\tilde{\zeta}_k$ constant).}
\label{fig1}
\end{figure}

In fact the equations are of generalized Lam\'e form in {\em Minkowski}
space, as studied in depth by Greene {\em et al} \cite{GKLS97} and Kaiser
\cite{dk2}.  The Floquet index $\mu_{\kappa}$ is shown in Fig.
(\ref{fig0}) as a function of $g^2/\lambda$ and $\kappa^2$. For
$g^2/\lambda = 2$ it reaches its maximum value of
$\mu_{\kappa} \simeq 0.238$ at $\kappa^2 = 0$, which coincides with the
global maximum of the Floquet index. Further, there exist an infinite
number of resonance bands in $g^2/\lambda$ for which
$\mu_{\kappa = 0} > 0$, showing that for large regions of parameter space
there is essentially exponential growth of super-Hubble metric
perturbations. This provides the first conclusive analytical proof of this
fact, confirming previous numerical work \cite{BKM,BTKM}.

In Fig. (\ref{fig1}) we plot $\tilde{\Phi}_k$, $\tQkp$ and $\tQkc$ for
$\kappa^2 = 10^{-20}$ and $g^2/\lambda = 2$. Note the perfect exponential
growth of $\tQkc$, while $\tQkp$ simply oscillates with constant
amplitude since it corresponds to $g^2/\lambda = 3$ for which there is
only a sub-Hubble resonance band. The Floquet index for $\tQkc$ is as
expected from the Lam\'e
analysis, while the Floquet index for $\tilde{\Phi}_k$ is much larger,
around $0.357$. This is expected from the constraint equation
(\ref{constraint2}) which shows that $\tilde{\Phi}'$ is sensitive to
the product $\tilde{\chi} \tQkc$, with {\em both} $\tilde{\chi}$ and
$\tQkc$ growing exponentially.

Expressed in dimensionless time and neglecting the $a'' \simeq 0$ term in
eq. (\ref{zeta}), $\tilde{\zeta}$  becomes
\beq
\tilde{\zeta}_k = \tilde{\Phi}_k + \sqrt{\lambda}\varphi_0 x
\tilde{\Phi}'_k\,,
\label{zeta2}
\eeq
which shows that $\tilde{\zeta}_k$ also grows exponentially, sharing the
same Floquet index as $\tilde{\Phi}_k$ before backreaction shuts off the
resonance. This is evident in the inset of Fig. (\ref{fig1}) where we
show $\tilde{\zeta}$ for $g^2/\lambda = 0$ (no growth) and $g^2/\lambda =
2$ (exponential growth).

Finally, note that the effective resonance parameter $q_{eff} =
g^2/\lambda$ can be as small as $q_{eff} = 2$ for the Floquet index to
reach its global maximum and for super-Hubble preheating to be strong.
This means that the $\chi$ field is not {\em necessarily} suppressed
during inflation \cite{supp}. This is discussed in detail in section
(\ref{init}). The suppression also does not apply in wide classes of
other preheating models \cite{BGMK}.

\subsubsection{Case 2: Backreaction included}

\begin{figure}
\epsfysize=2in
\epsffile{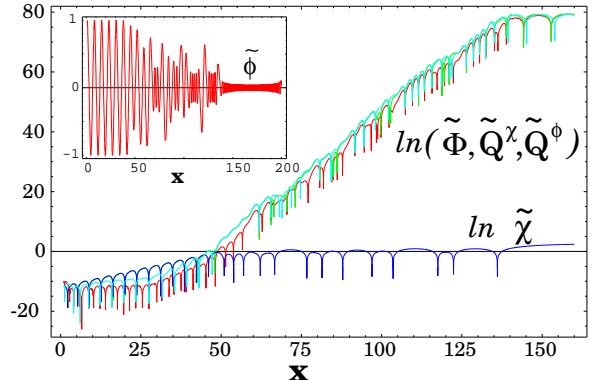}
\caption{The field $\tilde{\chi}$ and the gauge-invariant metric perturbations
$\tilde{\Phi}_k$, $\tQkc$ and $\tQkp$ for $\k^2 =10^{-20}$ and
$g^2/\lambda = 2$. Note the onset of backreaction in $\tilde{\chi}$ at $x
\simeq 50$ and the synchronisation in the metric perturbations for $x \geq
50$. {\bf Inset:}  The evolution of the inflaton condensate $\tvp$: note
the drop in amplitude and increase in oscillation frequency at $x \simeq
50$ and $135$. }
\label{fig2}
\end{figure}

We now consider the full equations (\ref{inflaton} -- \ref{qchi}) without
approximation. Fig. (\ref{fig2}) shows the results of simulations
including backreaction on the inflaton. For $x < 50$, the analysis in
{\em case 1}, presented above, applies. However, when $\ln \tilde{\chi}
\sim O(1)$, the backreaction in Eq. (\ref{inflaton}) becomes important,
changing the amplitude and frequency of $\tvp$ oscillations as shown in
the inset of Fig. (\ref{fig2}). This pushes $\tchi$ into a region of
narrow resonance with $\mu_{\tilde{\chi}} \simeq 0.05$.

Importantly, however, the metric perturbations do {\em not} stop growing
once $\chi$ moves into the region of narrow resonance. The couplings
between Eqs. (\ref{qphi}) and (\ref{qchi}) cause $\tQkp$ to begin growing
exponentially and {\em synchronisation} occurs: for $x >
50$ the perturbations $\tilde{\Phi}_k, \tQkc$ and $\tQkp$ all grow with
the {\em same}, large, Floquet index. This  shows that any analysis based
on the decoupled equations one finds  in the absence of metric
perturbations must be considered with caution, since it would (1)
underestimate the  Floquet indices and (2) incorrectly predict no growth
of $\tQkp$ and $\tdvp_k$ for $\kappa^2 \ll 1$.  From Eq. (\ref{smvar}) one
sees that the field fluctuations $\tdvp_k$ grow with the same, large
Floquet index as the metric perturbations.

Backreaction finally ends the growth of the metric perturbations
around $x \sim 135$ when the amplitude of $\tilde{\phi}$ drops and the
inflaton effective mass dramatically increases (see  Fig. \ref{fig2}).

\section{Preheating initial conditions}\label{init}

The key point that arises  from having resonant growth of the super-Hubble
metric modes with $\kappa \ll 1$  for $1 \leq g^2/\lambda  < 3$ is its
relation to initial conditions at the start of preheating. Convolving
these two will yield the post-preheating power spectrum.

The $\chi$ effective mass is given by
\beq
m_{\chi,eff}^2 = g^2 \phi^2\,.
\label{chimass}
\eeq
Ignoring the metric perturbations during inflation where they are small,
one can solve the Klein-Gordon equation in de Sitter spacetime for the
$\chi$ modes in terms of Hankel functions
\cite{SH99}:
\beq
\chi_k = c_{1}\sqrt{\eta} H^{(2)}_{\nu}(k\eta) + c_{2}\sqrt{\eta}
H^{(1)}_{\nu}(k\eta)\,,
\label{chi}
\eeq
where $c_1 = \sqrt{\pi}/2$, $c_2 = 0$ corresponds to the
adiabatic vacuum state and the order, $\nu$, of the Hankel function is
the root of
\beq
\nu^2 = \frac{9}{4} - \frac{m_{\chi}^2}{H^2} - 12\xi\,.
\label{nu}
\eeq
Here $\xi$ is the non-minimal coupling of the $\chi$ field to the
curvature, which we have put to zero throughout so far.
Asymptotically, as $k\eta \rightarrow 0$,  the Hankel functions are
\cite{dn}
\beq
H^{(1,2)}_{\nu}(k\eta) \rightarrow
\frac{i}{\pi}\Gamma(\nu)\left(\frac{k\eta}{2}\right)^{-\nu}\,,
\label{asymp}
\eeq
so that if ${\bf Re}(\nu) > 0$, the mode functions are infra-red
divergent. However, when ${\bf Re}(\nu) = 0$, it is simple to
show that the mode functions are finite as $k\eta \rightarrow 0$ and are
essentially $k$-independent.  Further, if ${\bf Re}(\nu) =
0$, the modes decay away as $a^{-3/2}$, that is,
exponentially in proper time. This is the origin of the claim that
preheating has no effect on super-Hubble metric perturbations \cite{supp}.

Now, in our case, since $\chi = 0$ minimises the potential at fixed
$\phi$, the $\chi$ field has no potential energy
during inflation and $H^2 \simeq  2\pi \lambda \phi^4/3$.
Hence when $\xi = 0$
\beq
\nu^2 = \frac{9}{4} - \frac{3 g^2}{2\pi \lambda \phi^2}\,,
\label{nu2}
\eeq
hence for $g^2/\lambda = 2$, $\nu^2 < 0$ when $\phi < 2/\sqrt{3\pi} \sim
0.7 M_{pl}$.  Now since preheating starts at around $\phi \sim 0.3
M_{pl}$,   $\nu^2 >
0$ during most of inflation.  This case differs
radically from the case $g^2/\lambda \gg 1$, where $\nu$ is complex many
e-foldings from the end of inflation. Hence, while the suppression of
$\chi$ modes is very strong when $g^2/\lambda \gg 1$\cite{supp}, for
$g^2/\lambda \in [1,3)$ there is little suppresion of cosmological $\chi$
modes.

Before discussing the post-preheating power spectrum we
make two points. Firstly, our potential, Eq. (\ref{2fieldpot}), is not
meant to be generic, and hence specific results
should be taken as proofs of principle rather than as generic. In
particular, the suppression of $\chi$ modes when $g^2/\lambda \gg 1$ is
absent in preheating with more general potentials \cite{BGMK}.

Secondly, the suppression that occurs for $g^2/\lambda \gg 1$ and
$\chi = 0$  (minimally coupling), does not {\em necessarily persist} when
$\xi \neq 0$. From Eq. (\ref{nu}) one sees that  $\nu^2 > 0$ is compatible
with  $m_{\chi}^2 \gg H^2$ (and hence little or no suppression) for large
negative couplings, $\xi$, to the curvature.

In fact, for sufficiently large negative $\xi$, $\nu > 3/2$ and the
resulting power spectrum has most power at {\em large} scales -- a red
tilt -- and field variances receive their dominant contribution from the
super-Hubble modes \cite{SH99}. Large non-minimal couplings in fact lead
to their own strong resonances due to the oscillation of the Ricci scalar
during preheating \cite{BL98,TMT}.

\section{The $\chi$ power spectrum}\label{power}

While our main aim in this paper has been to provide analytical evidence
for the existence of super-Hubble resonances, we turn here to the
question of the post-preheating power spectrum.

The power spectrum for an arbitrary field $X$ is defined by
\beq
P_X (k) \equiv \frac{k^3}{2\pi^2} \langle|X_k|^2\rangle \,,
\eeq
where $\langle\cdot\rangle$ denotes ensemble averaging (equivalently
volume averaging if $X$ is ergodic). From Eqs. (\ref{chi},\ref{asymp}),
the power-spectrum for $\chi$ at the end of inflation in the limit $k\eta
\rightarrow 0$  is thus given by:
\beq
P_{\chi}(k,\eta)  = \frac{k^3}{8\pi^3} \eta \Gamma^2(\nu)
\left(\frac{k\eta}{2}\right)^{-2\nu}\,.
\label{spec}
\eeq
This clarifies the situation immediately. For $\nu$ complex, the
power spectrum is steep $\propto k^3$. However, for $\nu$ real, the
power spectrum is closer to Harrison-Zel'dovich with spectral index
\beq
n = 3 - 2\nu\,,
\label{specindex}
\eeq
which tends to zero  as $\nu \rightarrow 3/2$ ($\phi \rightarrow \infty$).
Now it is true that $\nu$ becomes complex at the end of inflation, but
since we are interested in the $k \rightarrow 0$ part of the spectrum, the
appropriate value of $\nu$ is the one at early stages: $\phi \geq 2
M_{pl}$, for which $\nu \geq  \sqrt{2}$, leaving only a mild blue tilt.

The power spectrum (\ref{spec}) is then modified by preheating in a rather
trivial way: it is multiplied by a  factor
$\sim e^{2\mu \Delta x}$,  where $\Delta x$ is the time spent in
resonance. For $\kappa^2 \ll 1$, modes, this transfer function is almost
independent  of $k$ since the Floquet index only changes significantly
over $\Delta \kappa \sim O(1)$.

The above discussion paves the way for direct evaluation of the impact of
preheating on large scales. The obvious question remains: ``how much of
the $g^2/\lambda$ phase space will leave an imprint on the CMB?"  In this
model we may investigate this question analytically without relying on
numerical studies.

\subsection{Dividing up the $g^2/\lambda$ parameter space}

For cosmological implications we are typically only interested in
the $\kappa^2 \leq 10^{-50}$  section of the $\kappa^2-g^2/\lambda$
chart, Fig. (\ref{fig0}).  It is relatively easy to
show that the nodes of the resonance bands (where $\mu_{\kappa = 0}$
becomes non-zero) are given by $g^2/\lambda =
n(n+1)/2$, with $n$ natural. It is then easy to show that asymptotically
for large $g^2/\lambda$, half of all values of $g^2/\lambda$ lie in
resonance bands, and half lie in stable regions.

However, if $\xi = 0$, only for a much smaller measure of $g^2/\lambda$
values do the cosmological modes dominate the backreaction integral,
immediately ruling out those values of $g^2/\lambda$. This clinical
division of the $g^2/\lambda$ parameter space is due to the
conformal invariance of our model. For generic models
with massive fields the modes will move through the maxima of the
resonance bands and the issue of which couplings are ruled out is
more subtle, and probably needs to be done numerically.

On the other hand, modes near $g^2/\lambda = 1$ or $3$, will experience
super-Hubble growth and are also little suppressed during inflation.
However they  are unlikely to go nonlinear before backreaction
becomes important (since sub-Hubble modes grow the quickest - see Fig
\ref{fig0}). Whether these regions of the resonance band are ruled out is
therefore more subtle, but they are expected to leave an imprint on the
CMB.

For $g^2/\lambda \gg 1$ ($\xi = 0$), the $\chi$ spectrum is suppressed at
small $k$ and although super-Hubble modes experience resonant growth,
backreaction may well shut off the resonance before they become
important and affect the power spectrum in any significant way.
Equivalently $n \simeq 3$ in Eq. (\ref{specindex}) \cite{supp}.

The $g^2/\lambda$ parameter space is thus divided into four sub-regions:
(I) regions where backreaction comes from cosmological modes going
nonlinear. This is a small set in the conformal model and can be
ruled out directly. They form narrow regions around the maxima of the
Floquet index at $g^2/\lambda = 2, 8, ...$.  (II) regions with $0 <
\mu_{\k = 0} < \mu_{maximum}$, i.e. in the wings of the resonance
bands. The $Q_{\chi}$ variance is dominated by sub-Hubble modes but they
can nevertheless leave an imprint on the CMB if $g^2/\lambda < 10$. (III)
Regions with $\mu_{\k = 0} > 0$ but $g^2/\lambda > 10$ where suppression
during inflation is strong and little effect is seen. (IV) Stable bands
where $\mu_{k = 0} = 0$ and backreaction is irrelevant.


\section{$\chi$ self-interaction}\label{back}

We now consider the dual issue of the possible effects of the rapidly
growing metric perturbations on particle production.

\begin{figure}
\epsfysize=2in
\epsffile{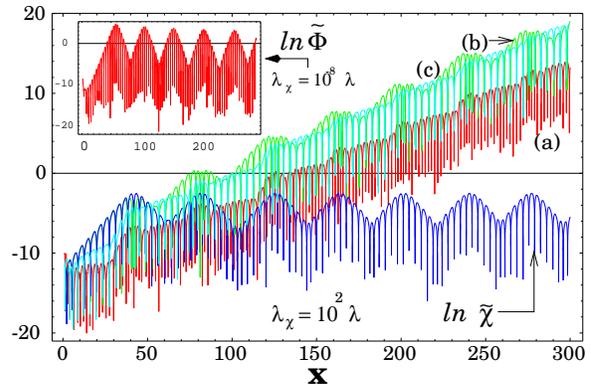}
\caption{Evolution of metric perturbations with
$\tilde{\chi}$ self-interaction: (a) $\tilde{\Phi}_k$ (b)
$\tQkc$ (c) $\tQkp$ and the $\tilde{\chi}$ field, for $\lambda_{\chi} =
10^{2}\lambda$, $\kappa^2 = 10^{-20}$ and $g^2/\lambda = 2$. While
self-interaction stops $\tilde{\chi}$ from growing, it fails to stop
exponential growth of the  metric perturbations. {\bf Inset:}
$\tilde{\Phi}_k$ evolution with all terms on the RHS of Eq's
(\ref{qphi},\ref{qchi}) set to
zero, and $\lambda_{\chi} = 10^{-8} \lambda$. The apparent shut-off of the
resonance is an artifact due to neglect of backreaction and coupling
between $\tQkp$ and $\tQkc$.}
\label{fig3}
\end{figure}

A  result from early studies of preheating, believed to be rather robust,
was that $\chi$ self-interaction lead to a rapid shut-off of the resonance
and inefficient decay of the inflaton \cite{selfint}. This is easy to
understand in the absence of metric perturbations since the
self-interaction acts like an effective mass $m_{\chi, eff}^2 \sim
3\lambda_{\chi} \langle \chi^2 \rangle$, which pushes  all modes out of
the dominant first resonance band and essentially ends the resonance. This
is evident in the inset of Fig. (\ref{fig3}) which shows $\tilde{\Phi}$ in
a model with $\lambda_{\chi} \chi^4/4$ self-interaction added to the
potential (\ref{2fieldpot}), with
$\lambda_{\chi} = 10^{-8} \lambda$ for {\em case 1}, i.e.
the equations (\ref{qphi},\ref{qchi}) with all terms on the RHS's set to
zero and $\tvp = cn(x,1/\sqrt{2})$.  Even with this tiny
self-interaction, backreaction appears to be  extremely efficient at
ending the resonant growth of super-Hubble fluctuations.

This, however, is an artifact of neglecting metric perturbations which
manifests itself as the couplings between the equations (\ref{qphi},
\ref{qchi}).  The main
graphs in Fig. (\ref{fig3}) show a very different behaviour, even when  a
much larger self-interaction, $\lambda_{\chi} = 10^2 \lambda$, is used to
emphasise the discrepancies with the old picture. The metric and field
fluctuations continue to grow exponentially long after the $\tilde{\chi}$
field has ended its growth. How is  this to be understood ?

The growth of $\chi$ forces it out of
the resonance band and causes it to stop growing before it has a chance to
affect the inflaton evolution through Eq. (\ref{inflaton}). This is in
complete contrast to the case in
Fig. (\ref{fig2}). Since it is the amplitude and frequency of inflaton
oscillations that predominantly controls the resonant growth of field and
metric fluctuations, the large $\chi$ self-interaction actually aids
the growth of metric perturbations in the sense that it removes the
means of backreaction on the inflaton condensate \footnote{Using
$\langle\tilde{\chi}^2\rangle$ instead of $\tilde{\chi}^2$ in Eq.
(\ref{inflaton}) will cause the field fluctuations in Fig. (\ref{fig3})
to stop growing at some large, but finite, value. However, our main point
is to stress the new channels through which the resonance can
proceed when metric perturbations are included.}.

This provides an elegant example of both the qualitative and quantitative
importance of including metric perturbations in preheating. This is true
not only when one is interested in understanding the spectrum of
perturbations for the CMB, but also when calculating relic abundances of
particles produced during preheating. Relevant examples are the abundances
of gravitinos, GUT bosons and super-heavy dark matter whose decays may
contribute to the observed ultra-high energy cosmic ray flux \cite{app}.

\section{Conclusions}

In conclusion, we have attempted to convince the reader
of a number of controversial possibilities: (1)
gauge-invariant, super-Hubble scalar metric perturbations can grow
exponentially,  modulo power-law behaviour,  in an {\em expanding}
universe. We did this by reducing the gauge-invariant equations to
generalised Lam\'e form. (2) The quantity $\zeta$, often used to transfer
the power spectrum of fluctuations to decoupling by virtue of its
constancy, may be  useless in general during preheating due to its
exponential amplification. (3) Metric perturbations can significantly
alter estimates for the maximum variances of field fluctuations needed for
accurate prediction of relic particle abundances.

This last point is particularly transparent  when the inflaton decay
products exhibit strong self-interaction. In the
absence of metric perturbations the self-interaction stops preheating
almost completely \cite{selfint}. When metric perturbations are included
however, the resonances proceed  through new channels and can be strong
even at large self-interaction, greatly enhancing final field variances
and number densities.  These issues will be discussed in depth in
\cite{VB}.

The issue of the post-preheating $\chi$ power spectrum has been discussed
as a function of the coupling $g^2/\lambda$. We show that while there
are an infinite number of super-Hubble resonance bands, only for a small
range of values,  $g^2/\lambda < 10$,  can one expect modifications of
standard inflationary predictions for cosmology since these values are
partially or completely protected from strong suppression during
inflation. In more general, non-conformal, models the situation will be
more complex as modes are pulled through resonance bands by the
expansion. Crucially the mode suppression mechanism may also be absent
\cite{BGMK}.

To end we discuss an issue little studied to date. Eliminating all other
perturbation variables, the $\Phi$ equation of motion typically has
singular coefficients $\propto \dot{\phi}^{-1}$ \cite{MFB,KS84}. Indeed,
this was the main original motivation for using the Sasaki-Mukhanov
variables \cite{early2,later}. Insight into $\Phi$ evolution is provided
by the duality
between the Klein-Gordon and 1-d spatial Schr\"odinger equations exploited
in \cite{b98}. In this approach the resonance band structure in
$k$-space becomes dual to the spectrum of the corresponding  Schr\"odinger
equation. Now consider the dimensionless Scarf Hamiltonian:
\beq
H \psi \equiv \(-\frac{d^2}{dz^2} + \frac{A}{\sin^2 z}\)\psi(z) = \lambda
\psi
\label{scarf}
\eeq
whose potential is singular periodically when $\sin z = 0$. This
Hamiltonian is known to have a spectrum with band structure for $-1/4 < A
< 0$ and discrete eigenvalues otherwise \cite{scarf}. On the complement of the
spectrum, ${\bf R}-\sigma(H)$, the corresponding Floquet index is real
and positive \cite{b98}.  This explains why the
behaviour of $\tilde{\Phi}_k$ can be  qualitatively  similar to that of
$\tQkc$ and $\tQkp$, which obey {\em non-singular} equations.

We thank Kristin Burgess and David Kaiser for detailed, insightful
comments \& Rob Lopez, Roy Maartens, Levon Pogosian and David Wands for
useful discussions.  FV acknowledges support from CONACYT scholarship
Ref:115625/116522.


\end{document}